\documentclass[ amsmath,amssymb,aip,reprint,superscriptaddress]{revtex4-2}

\usepackage{graphicx,mathtools,subcaption}
\usepackage{appendix,xcolor}
\graphicspath{{figs/}}
\usepackage{lineno}

\newcommand{\crea}{\hat{a}^\dag}
\newcommand{\anni}{\hat{a}}
\newcommand{\im}{\mathrm{i}}

\newcommand{\te}[1]{\mathrm{#1}}
\newcommand{\expo}[1]{\mathrm{exp}\left\{ #1 \right\}}
\newcommand{\ex}[1]{\mathrm{e}^{#1}}
\newcommand{\bc}[1]{\textcolor{black}{#1}}
\newcommand{\rc}[1]{\textcolor{black}{#1}}
\newcommand{\braket}[1]{\langle #1 \rangle}

\newcommand{\matrixBlock}[4]{\begin{bmatrix} #1 & #2 \\ #3 & #4 \end{bmatrix}}
\usepackage[lined,ruled,linesnumbered]{algorithm2e}

\begin{document}
%\linenumbers
%\preprint{}

\title{Effect of partial distinguishability on quantum supremacy in \bc{Gaussian boson sampling}}


\author{Junheng Shi}
%\thanks{}
\affiliation{State Key Laboratory of Precision Spectroscopy, School of Physical and Material Sciences, East China Normal University, Shanghai 200062, China}
\affiliation{New York University Shanghai, 1555 Century Ave, Pudong, Shanghai 200122, China}
%\altaffiliation{1State Key Laboratory of Precision Spectroscopy, School of Physical and Material Sciences, East China Normal University, Shanghai 200062, China}
\author{Tim Byrnes}
\email[]{corresponding author: tim.byrnes@nyu.edu}
\affiliation{New York University Shanghai, 1555 Century Ave, Pudong, Shanghai 200122, China}
\affiliation{NYU-ECNU Institute of Physics at NYU Shanghai, 3663 Zhongshan Road North, Shanghai 200062, China}
\affiliation{State Key Laboratory of Precision Spectroscopy, East China Normal University, Shanghai 200062, China}
\affiliation{National Institute of Informatics, 2-1-2 Hitotsubashi, Chiyoda-ku, Tokyo 101-8430, Japan}
\affiliation{Department of Physics, New York University, New York, New York 10003, USA}
%\thanks{corresponding author: tim.byrnes@nyu.edu}


%\date{May 20,2021}

\begin{abstract}
Gaussian boson sampling (GBS) allows for a way to demonstrate quantum supremacy with the relatively modest experimental resources of squeezed light sources, linear optics, and photon detection. In a realistic experimental setting, numerous effects can modify the complexity of the sampling, in particular loss, partial distinguishability of the photons, and the use of threshold detectors rather than photon counting detectors. In this paper, we investigate GBS with partial distinguishability using an approach based on virtual modes and indistinguishability efficiency. We develop a model using these concepts and derive the probabilities of measuring a specific output pattern from partially distinguishable and lossy GBS for both types of detectors. In the case of threshold detectors, the probability as calculated by the Torontonian is a special case under our framework. By analyzing the expressions of these probabilities \rc{we propose an efficient auxiliary classical simulation algorithm which can be used to calculate the probabilities. Our model and the auxiliary algorithm provide foundations for an approximation method for the probability calculation and simulation algorithms not only compatible with existing state-of-the-art simulation algorithms for ideal GBS but also reduce their complexities exponentially depending on the indistinguishability. Using approximation method and the simulation algorithms we show how the boundary of quantum supremacy in GBS can be pushed further by partial distinguishability.}  % insert abstract here
\end{abstract}

%\keywords{}

\maketitle

\section*{Introduction}
%\subsection*{\label{}}
%\subsection{}
%\subsubsection{}

It is commonly believed that quantum computers have the ability to outperform classical computers and falsify the Extended Church-Turing Thesis \cite{aaronson_computational_2013}. In order to demonstrate this quantum speedup, one requires not only a large number of qubits but also sufficiently suppressed logical errors. Decoherence typically removes the quantum aspect of the computation, such that it becomes possible to classically simulate the computation efficiently. Specifically, quantum supremacy is claimed when the quantum device achieves a superpolynomial quantum speedup. This was the case in the comparison between the quantum processor created by Martinis and co-workers \cite{arute_quantum_2019} and the \bc{ classical} simulation algorithms of Ref. \cite{pednault_leveraging_2019} and recently from Ref. \cite{pan_simulating_2021}.

Another heated battlefield in the context of quantum supremacy is the boson sampling (BS) problem where a passive linear optics interferometer, Fock states at the input, and photon-detectors are placed at the output. It was proved by Aaronson \& Arkhipov \cite{aaronson_computational_2013} that the output distribution of such device cannot be efficiently simulated by a classical computer in polynomial time \cite{valiant_complexity_1979,scheel_permanents_2004} because it corresponds to the permanent of a given transformation matrix. However, challenges arise in the experimental implementation Aaronson \& Arkhipov's boson sampling (AABS). One of them is the low generation efficiency of single photons. To truly outperform the state-of-art classical simulation algorithm for AABS \cite{neville_classical_2017}, at least 50 photons are required at the input. But due to the low single photon efficiency of solid state sources \cite{wang_high-efficiency_2017,he_time-bin-encoded_2017,loredo_boson_2017} which is typically between 0.2 and 0.3, the largest experimentally demonstrated number of single photon inputs in AABS has only been 20 to date \cite{wang_boson_2019}. It was then proposed that this challenge can be tackled by changing the input light from Fock states to Gaussian states which was first conceived in the form of Scattershot Boson Sampling (SBS) \cite{lund_boson_2014,barkhofen_driven_2017,you2017multiparameter}  and later developed to GBS \cite{hamilton_gaussian_2017,kruse_detailed_2019}. The probability of measuring a specific output pattern in GBS corresponds to the Hafnian of a matrix \cite{caianiello_eduardo_r_combinatorics_1973} which has been also proven as being computationally hard \cite{valiant_complexity_1979}. While generally the significance of BS is in relation to quantum supremacy, GBS has also potential applications in finding subgraphs \cite{arrazola_using_2018} and molecular vibronic spectra \cite{huh_boson_2015}. 

GBS was first realized in small-scale experiments \cite{zhong_experimental_2019,paesani_generation_2019}. \bc{In 2020 quantum supremacy was claimed by Pan, Lu, and co-workers in a GBS device named \textit{Jiuzhang}} \cite{zhong_quantum_2020} which was able to generate up to 76 photon detection events at the output of a 100-mode interferometer. It should be noted the average output photon detection events is in the vicinity of 40 which means that it was not in the collision-free regime which most of the current GBS theories fall into. \bc{Recently, the same group released an improved version with lower losses \cite{zhong_phase-programmable_2021}, which generates 113-photon detection events in a 144-mode circuit, going further away from the collision-free regime.}  Another major aspect of the experiment is the type of detection device used. In the experiment, threshold detectors, rather than photon number resolving (PNR) detectors were used, which only indicate the presence or lack of a photon. The use of threshold detectors changes the underlying theory because the output probability distribution is no longer determined by a Hafnian, but the Torontonian of the matrix representing the linear optical network \cite{quesada_gaussian_2018,li_benchmarking_2020}.

\bc{The Hafnian and the Torontonian have the same computational complexity for the ideal case, which is $O(N^32^N)$ for a $2N\times2N$ matrix. Despite having the same complexity, their origins are in fact rather different. In the case of the Hafnian, the complexity arises due to the large number of permutations of matrix elements and the lack an efficient algorithm (i.e. which is available for other matrix functions such as the determinant).  Meanwhile, the complexity of computing the Torontonian comes from the computation of $2^N$ determinants. Recently, it was shown that a quadratic speedup for simulating GBS can be achieved \cite{bjorklund_faster_2019}, based on  the observation that the probability of a $N$-photon detection for a pure Gaussian state only scales as $2^{\frac{N}{2}}$. This allows for the reduction in complexity of simulating GBS with PNR detectors \cite{quesada_quadratic_2021} and with threshold detectors \cite{bulmer_boundary_2021} even for the case of lossy GBS with the help of the Williamson decomposition and loop Hafnian. It however is not as effective in the case of mixed Gaussian states since it requires calculating a large number of loop Hafnians associated with the pure state decomposition of the mixed state.}
%These two matrix functions have the same computational complexity for the ideal case which is $O(N^32^N)$ for a $2N\times2N$ matrix. Despite this, the origin of the complexity is rather different for the two cases. In the case of the Hafnian, the complexity arises due to the large number of permutations of matrix elements; meanwhile the complexity of computing Torontonian comes from the computation of $2^N$ determinants.

Another major challenge is to understand the impact of experimental imperfections such as losses \cite{rahimi-keshari_sufficient_2016, oszmaniec_classical_2018}, network noise \cite{shchesnovich_noise_2019} and partial photon distinguishability \cite{rohde_optical_2012,tichy_stringent_2014,shchesnovich_partial_2015, tichy_sampling_2015, renema_efficient_2018}. Similarly to AABS, the original theory of GBS only handles the ideal case and is not easily extendable to include these imperfections. So far, only losses and detector dark counts in GBS have been analyzed \cite{qi_regimes_2020}. Ref. \cite{qi_regimes_2020} is based on the framework of Ref. \cite{rahimi-keshari_sufficient_2016} where the generalized Wigner function is used to demarcate the region of efficient classical simulation as a function of losses and detector dark counts. In contrast, no analysis has been carried out for partial photon distinguishability in GBS, except for a recent paper \cite{renema_simulability_2020} claiming to extend their results for partial distinguishability in AABS to that in GBS. The results of Ref. \cite{renema_simulability_2020} are somewhat unsatisfactory in the sense that their results for GBS must be applied in a context similar to AABS. For example, their investigation is strictly limited to the collision-free regime with weak squeezing, and they presume that the number of photons generated by the source is fixed and known at the beginning of the derivation. This is a rather strong assumption in relation to the indeterminate photon number nature of Gaussian states, especially in the presence of loss. In addition, Ref. \cite{renema_simulability_2020} only discusses GBS with PNR detectors, such that the theory is not applicable to the current GBS experiments using threshold detectors. 

In this paper, we provide an investigation of the partial distinguishability problem in GBS, not limited by the collision-free or determinate photon number assumptions. We develop a model for partial distinguishability and apply it to GBS with both PNR and threshold detectors. In this model we introduce a new parameter called the indistinguishability efficiency. Along with other existing parameters such as the squeezing parameter of the input light and transmission rate introduced from the lossy GBS model, it forms a composite parameter that affects the probability distribution and its underlying structure, similar to how transmission rate and dark counts work together to determine the classical simulatability of imperfect AABS \cite{rahimi-keshari_sufficient_2016}. We define virtual modes to incorporate the distinguishable photons that do not interfere with other photons but contribute to the photon detection at the end. \bc{We note that a similar approach was also used in  Ref.\cite{thomas_general_2021} where it was used for investigating the trade-off between Hong-Ou-Mandel interference visibility and photon generation efficiency for heralded single photon source. } For GBS with PNR detectors, the resulting probability is calculated as a sum of all possible combinations of different outputs of these states. Despite starting from completely different models, we find our characterization of the distinguishable photons in GBS matches perfectly with that derived for AABS \cite{aaronson_bosonsampling_2014}, which adds evidence towards the validity of our model. For GBS with threshold detectors, we include both partial distinguishability and losses in deriving the expression for the probability. In order to include the distinguishable photons from the virtual modes, we abandon the commonly employed Torontonian method and propose a method based on the marginal probability and prove that the probability defined by Torontonian is a special case of our result. 

We finally discuss how partial distinguishability affects quantum supremacy in light of our results. \rc{In order to give a comprehensive analysis of its effect, we investigate both regimes with more focus on the former one: the probability calculation of certain output pattern and the generation of one sample.} For partially distinguishable GBS with PNR detectors, since every term in the output probability corresponds to a particular number of indistinguishable photons, this determines the computational cost of calculating that term. By showing that the cost increases exponentially with the number of indistinguishable photons, we obtain an efficient approximation scheme by considering only a fraction of all terms which involves a smaller number of indistinguishable photons. In other words, GBS becomes more “classical" with reduced indistinguishability. We check the fidelity and complexity of the approximation which depends mainly on the indistinguishability efficiency. The fidelity of our approximation increases exponentially with decreasing indistinguishability. \rc{We also propose two simulation algorithms, one for PNR detector and one for threshold detector based on our model and proposed efficient classical algorithm. With these algorithms, the complexities of the state-of-the-art simulation algorithms\cite{quesada_quadratic_2021,bulmer_boundary_2021} can be even further reduced exponentially depending on the indistinguishability efficiency.}  In this way, we show that partial distinguishability affects quantum supremacy.

\section*{Results}

\subsection*{The model of partial distinguishability}

A typical GBS scheme consists of three parts: an interferometer with $K$ spatial ports for inputting photons and $K$ ports for outputs; $M$ of these input ports are fed with squeezed vacuum states and each output port has a detector. We note that what we refer to as “ports" are commonly referred to as “modes" in most of the literature on boson sampling since all photons are typically assumed to be indistinguishable. In this paper, we use the word “port" since each port may consist of multiple modes. The interferometer is characterized by a $K \times K$ Haar random matrix \text{T} where its columns correspond to the input ports and its rows correspond to the output ports.

It is convenient to represent a Gaussian state by a quasiprobability distribution (QPD) because the first two statistical moments of the QPD --- the displacement vector and covariance matrix --- are sufficient to fully characterize the density matrix \cite{ferraro_gaussian_2005,weedbrook_gaussian_2012}. Since the Gaussian states we are dealing with always have zero displacement vector, we write $\boldsymbol{\rho} = \boldsymbol{\rho}(\mathbf{V})$ and only use the covariance matrix \textbf{V} to represent the density matrix. Its definition is given by 
\begin{equation}
    V_{kl} = \frac{1}{2} \braket{\{\Delta\hat{\mathrm{x}}_k, \Delta\hat{\mathrm{x}}_l\}},
\end{equation}
where $\Delta\hat{\mathbf{x}} = \hat{\mathbf{x}} - \langle \hat{\mathbf{x}} \rangle $ and the quadrature field operators are defined as $\hat{\mathbf{x}} = [\hat{q}_1, \hat{p}_1, ..., \hat{q}_K, \hat{p}_K], \hat{q}_k = \anni_k + \crea_k, \hat{p}_k = \im(\crea_k - \anni_k)$, and $\hat{a}_k$ are the annihilation operators for the $k$th port. We let $k \in [1,K]$ throughout this paper.

By directly using the existing model for lossy GBS as in Ref. \cite{qi_regimes_2020}, the covariance matrix of squeezed vacuum inputs with losses is 
    \begin{equation}
        \begin{split}
            &\mathbf{V}^{(0)} =
            \\
            &\bigoplus_{m=1}^M
        \begin{bmatrix}
            \eta_\mathrm{t}\ \ex{2r_m} + 1 - \eta_\mathrm{t} & 0
            \\
            0 &\eta_\mathrm{t}\ \ex{-2r_m} + 1 - \eta_\mathrm{t}
        \end{bmatrix}
        \bigoplus
        \textbf{I}_{2K-2M}\label{InputCV},
        \end{split}
    \end{equation}
where $ \eta_\mathrm{t} $ is the overall transmission rate:
\begin{equation}
    \eta_\text{t} = \eta_\mathrm{s}\eta_\mathrm{u}\eta_\mathrm{d}.
\end{equation}
Here, $\eta_\mathrm{s}$ is the transmission of the inputs (sources) before entering the interferometer, $ \eta_\mathrm{u} $ is the transmission for the uniform loss inside the interferometer, and $\eta_\mathrm{d}$ is the detection efficiency. In principle, we should only include $\eta_\mathrm{s}$ in Eq.(\ref{InputCV}) because it describes the Gaussian state before entering the interferometer, but since our model is compatible with the result in Ref. \cite{qi_regimes_2020,neville_classical_2017}, we combine $\eta_\mathrm{s}$, $\eta_\mathrm{u}$ and $\eta_\mathrm{d}$ at this stage. Here, $\mathbf{I}_{n}$ is the $ n \times n $ identity matrix which is the covariance matrix of the vacuum state. A standard assumption that is typically used in experiments such as Ref. \cite{zhong_quantum_2020} is that all inputs have identical squeezing parameter $r_m = r$. 

    \begin{figure}[t]
        \includegraphics[width=0.5\textwidth]{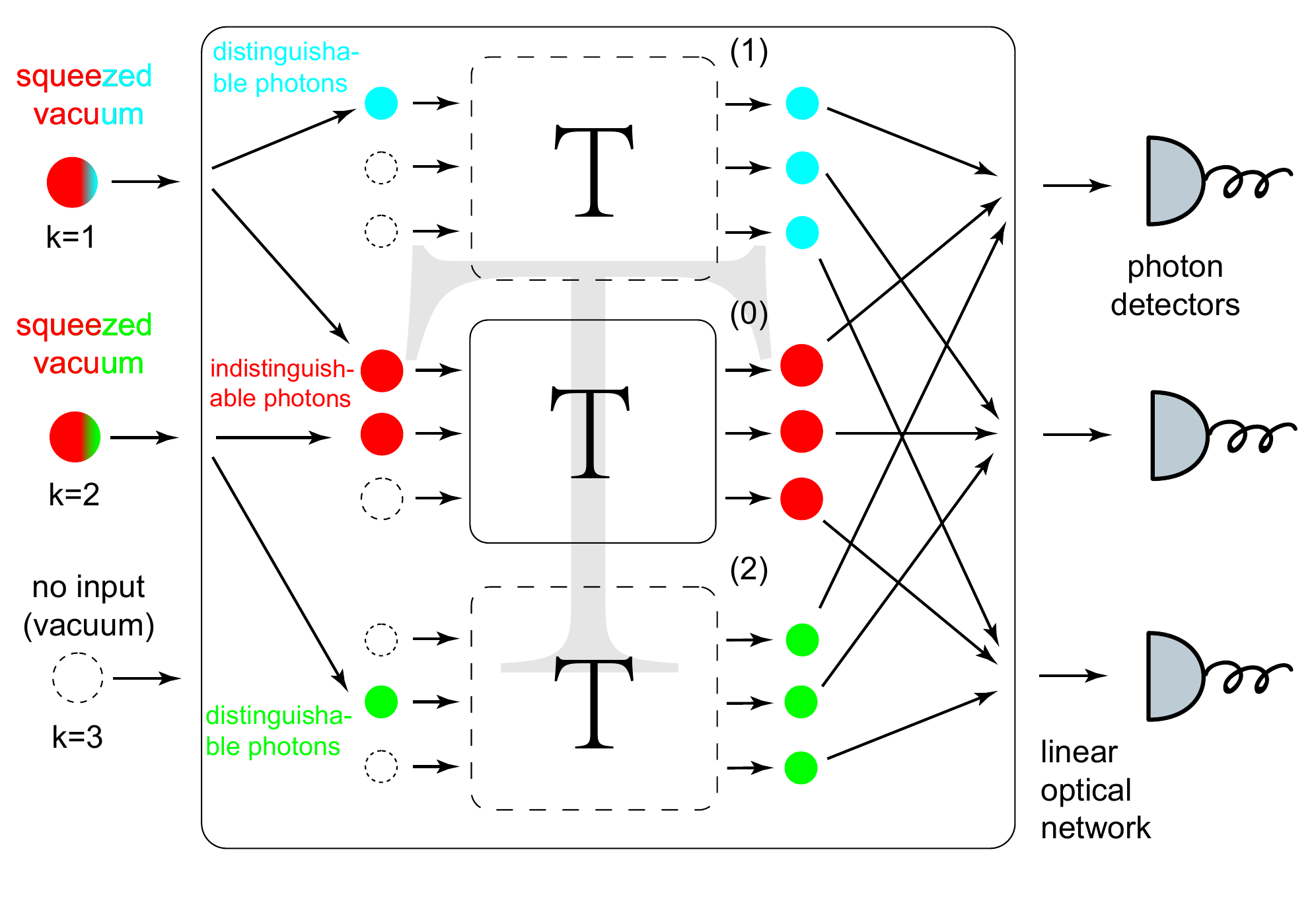}% Here is how to import EPS art
        \caption{\label{fig:model} \raggedright A pictorial representation of our model. For the simplicity of illustration, we show a $3\times 3$ interferometer ($K$=3) characterized by \textbf{T} with input light at first two ports ($M$=2). When partially distinguishable input light (represented by the mixed color ball for the input ports) enters the interferometer, they decompose into non-interfering modes and propagate inside the interferometer independently until they leave the interferomoter and collectively meet the detectors at the output ports. The thin solid-lined rectangle represents the interferometer for the indistinguishable photons. The dashed rectangles represent the interferometer for the distinguishable virtual modes. }
    \end{figure}

For GBS, partial distinguishability originates from imperfect input light where minor shifts in time or frequency is the origin of the distinguishability \cite{tichy_sampling_2015,renema_efficient_2018}. To characterize it, in our model all input light is initially indistinguishable. Before entering the interferometer, \bc{the photons go through a process where they have} the probability of becoming a distinguishable photon. These photons do not interfere with other photons during their propagation in the interferometer, but will be registered by the detectors at the output. In order to satisfy this assumption, we create an additional virtual mode for every port which has input light. The Gaussian state in each virtual mode is characterized as $ \boldsymbol{\rho}^{(m)}(\mathbf{V}^{(m)})$. Throughout this paper, the superscript $(m)$ is used to denote the quantities and parameters related to the distinguishable photons in the $m$th virtual mode and we let $m \in [1,M]$. \bc{The superscript ${(0)}$ will be reserved for the indistinguishable mode. We will also use superscript ${(m')}$ with $m' \in [0,M]$ when including both indistinguishable and distinguishable cases.} These virtual modes are initially in vacuum states, $\mathbf{V}^{(m)} = \mathbf{I}_{2K}$, until they are fed with photons through the corresponding port from the indistinguishable mode. The transformation of photons from the indistinguishable mode to the virtual mode is characterized by the unitary operator:
\begin{equation}
    U_m = \expo{ \theta ( \hat{a}_m^{\dag}\hat{b}_m^{(m)} - { \hat{b}_m^{(m)\dag} } \hat{a}_m )}, \label{unitaryOp}
\end{equation}
where $\hat{a}_m^{\dag}$ and $\hat{a}_m$ are the creation and annihilation operators of the $m$th port of the indistinguishable mode, $\hat{b}_m^{(m)\dag}$ and $\hat{b}_m^{(m)}$ are the creation and annihilation operators of the $m$th port of the $m$th virtual mode. Here, the subscript $m$ is to indicate which port. We define the indistinguishability efficiency as the probability of a photon not exchanged from the indistinguishable mode to the virtual modes, denoted as $\eta_\mathrm{ind}$. Under the effect of Eq.(\ref{unitaryOp}), it satisfies the relation:
\begin{equation}
    \eta_{\mathrm{ind}} = \cos^2\theta .
\end{equation}
There is some similarity between this model and that for lossy GBS \cite{rahimi-keshari_sufficient_2016,qi_regimes_2020}, but the major difference is that the photons in the virtual modes are not lost, instead, they go on to propagate in the interferometer until they are detected at the output.  
%We presume that such process occurs before photons enter the interferometer because most of the partial distinguishability comes from the photon sources. Nevertheless, similar to the model for lossy boson sampling, such presumption already incorporate the partial distinguishability occurs after the interferometer. 
%\bc{ This is mainly because that for a theoretical model where the interferometer is fully accounted by one single transformation matrix, there is no interference between photons starting from the same mode. }
%(Also, because the indistinguishability is rather high in modern experiment, the average photon number of one single virtual mode is close to ?, as we show later in Appendix ?)) 

Now we obtain the new density matrices for all $M$+1 modes after applying the distinguishable-indistinguishable transformation by taking the partial trace:
\begin{align}
    &\widetilde{\boldsymbol{\rho}} = \left( \prod_{m=1}^M U_m \right) \bigotimes_{m'=0}^M \boldsymbol{\rho}^{(m')} \left( \prod_{m=1}^M U_m \right)^\dag 
    \\
    &\widetilde{\boldsymbol{\rho}}^{(m')} = \mathrm{tr}_{m'}(\widetilde{\boldsymbol{\rho}}).
\end{align}
Since we use the covariance matrix to represent the density matrix, we give the  covariance matrices of all $M$+1 modes:
\begin{align}
    &\widetilde{\mathbf{V}}^{(0)} = \bigoplus_{m=1}^M
        \begin{bmatrix}
            \mathcal{X}_\te{ind} & 0
            \\
            0 &\mathcal{Y}_\te{ind}
        \end{bmatrix}
        \bigoplus
        \textbf{I}_{2K-2M} \label{CVindis}
    \\
    &\widetilde{\mathbf{V}}^{(m)} = \textbf{I}_{2m-2} \bigoplus
        \begin{bmatrix}
            \mathcal{X}_\te{dis} & 0
            \\
            0 & \mathcal{Y}_\te{dis}
        \end{bmatrix}
        \bigoplus
        \textbf{I}_{2K-2m}, \label{CVdis}
\end{align}
where we have 
\begin{align}
    &\mathcal{X}_\te{ind} = \eta_\mathrm{t}\eta_\mathrm{ind}\ \ex{2r} + 1 - \eta_\mathrm{t}\eta_\mathrm{ind}
    \\
    &\mathcal{Y}_\te{ind} = \eta_\mathrm{t}\eta_\mathrm{ind}\ \ex{-2r} + 1 - \eta_\mathrm{t}\eta_\mathrm{ind}
    \\
    &\mathcal{X}_\te{dis} = \eta_\mathrm{t} (1-\eta_\mathrm{ind})\ \ex{2r} + 1 - \eta_\mathrm{t}(1-\eta_\mathrm{ind})
    \\
    &\mathcal{Y}_\te{dis} = \eta_\mathrm{t}(1-\eta_\mathrm{ind})\ \ex{-2r} + 1 - \eta_\mathrm{t}(1-\eta_\mathrm{ind}).
\end{align}

    In our model, the output pattern measured at detection does not only come from the indistinguishable mode. It is the combined output pattern of all $M$+1 modes. Consider the case of ideal PNR detection. If the output pattern of the $m'$th mode is $\mathbf{s}^{(m')} = [s_1^{(m')},s_2^{(m')},...,s_K^{(m')}]$ where $s_k^{(m')}$ denotes the number of output photons at $k$th port of $m'$th mode, then the overall output pattern is 
    \begin{align}
        \mathbf{s} = \sum_{m'=0}^M \mathbf{s}^{(m')}, \label{output_pattern_1}
    \end{align}
    and for each port we have
    \begin{equation}
        s_k = \sum_{m'=0}^M s_k^{(m')}. \label{output_pattern_2}
    \end{equation}

\subsection*{Partially distinguishable GBS with PNR detectors \label{sec:PNR} }

    We now proceed to calculate the probability distribution of the output for PNR detectors.
    For the original GBS with perfect indistinguishability, the probability of observing an output pattern $\textbf{s} = [s_1, s_2, ..., s_K]$ where $s_k$ is the number of detected photons at the output of the $k$th port is given by the equation:
    \begin{equation}
    P(\mathbf{s}) = 
    \dfrac{\mathrm{Haf}(\mathbf{A}_\mathbf{s})}{s_1!...s_K!\sqrt{\det(\mathbf{Q})}}, \label{oriHaf} 
    \end{equation}
    where Haf($\cdot$) stands for the matrix function Hafnian and $\mathbf{A}_\mathbf{s}$ is the selected kernel matrix which is obtained by repeating rows and columns of the kernel matrix according to \textbf{s}. The kernel matrix is constructed out of the covariance matrix. \textbf{Q} is the covariance matrix of Q-function \cite{barnett_methods_2002}.
    
    For partially distinguishable GBS under our model, we need to first decompose the overall output pattern into different combinations of $M$+1 patterns, namely $\textbf{s}^{(0)}$, $\textbf{s}^{(1)}$ ... $\textbf{s}^{(M)}$, according to Eq.(\ref{output_pattern_1}) and Eq.(\ref{output_pattern_2}). We define $P^{(m')}(\mathbf{s}^{(m')})$ as the probability to obtain output pattern $\mathbf{s}^{(m')}$ for the $m'$th mode. The probabilities of all possible combinations are then combined to obtain the overall probability:
    \begin{equation}
        P(\textbf{s}) = \sum_{ \substack{\textbf{s}^{(0)},\textbf{s}^{(1)},...,\textbf{s}^{(M)}\\ \{ \sum_{m'=0}^M \mathbf{s}^{(m')}=\mathbf{s}\}}}
        \prod_{m'=0}^M P^{(m')}(\textbf{s}^{(m')}). \label{ProbPNRtotal}
    \end{equation}
    
    In Eq.(\ref{ProbPNRtotal}) there are a total of $\prod_{k=1}^K (s_k^{(0)}+1)$ possible configurations of $\mathbf{s}^{(0)}$, ranging from $[0,...,0]$ to $[s_1,...,s_K]$. Now we regroup them by the total number of photons in a configuration, denoted as $N^{(0)}$:
    \begin{equation}
        N^{(0)} = \sum_{k=1}^K s_k^{(0)},
    \end{equation}
    and rewrite Eq.(\ref{ProbPNRtotal}) as
    \begin{equation}
        P(\textbf{s}) = \sum_{n=0}^N \sum_{\substack{\mathbf{s}^{(0)}\\\{N^{(0)}=n\}}} 
        P^{(0)}(\mathbf{s}^{(0)})
        P_{\mathrm{dis}}(\mathbf{s}-\mathbf{s}^{(0)}), \label{regrouped_ProbPNR}
    \end{equation}
    where
    \begin{equation}
        P_{\mathrm{dis}}(\mathbf{s}_{\mathrm{dis}}) = \sum_{ \substack{\textbf{s}^{(1)},...,\textbf{s}^{(M)}\\ \{ \sum_{m=1}^M \mathbf{s}^{(m)}=\mathbf{s}_{\mathrm{dis}}\}}} \prod_{m=1}^M P^{(m)}(\textbf{s}^{(m)}), \label{total_P_dis}
    \end{equation}
    where $N$ is the total number of photons of the overall output pattern:
    \begin{equation}
        N = \sum_{k=1}^K s_k.
    \end{equation}
    Here, we consider the photons from all the virtual modes as a whole because later we propose an classical sampling algorithm for their combined probability in Eq.(\ref{total_P_dis}).
    
    In the Methods section, we obtain the covariance matrix  \textbf{Q} and kernel matrix \textbf{A} of all $M$+1 modes. Applying them to Eq.(\ref{oriHaf}) we obtain the specific probability distribution for each state. The expression for the probability with respect to the indistinguishable mode takes the same form as Eq.(\ref{oriHaf}):
    \begin{equation}
        P^{(0)}(\mathbf{s}^{(0)}) = 
        \dfrac{\mathrm{Haf}(\mathbf{A}_\mathbf{s}^{(0)})}{s_1^{(0)}!...s_K^{(0)}!\sqrt{\det(\mathbf{Q}^{(0)})}}. \label{indisHaf}
    \end{equation}
    For the distinguishable modes, on the other hand, the change in the form of kernel matrix leads to the reduction of the Hafnian matrix function to simple multiplication:
    
    \begin{align}
        &P^{(m)}(\mathbf{s}^{(m)}) = 
        \dfrac{  G(N^{(m)})  }{\sqrt{\det(\mathbf{Q}^{(m)})}}
        \prod_{k=1}^K \dfrac{|T_{k,m}|^{2s_k^{(m)}}}{s_k^{(m)}!}, \label{P_dis}
        \\
        &G(N^{(m)}) = \sum_q \frac{ (N^{(m)}! )^2 }{(q!!)^2 (N^{(m)} - q)! } \ \beta_\te{d}'^q\ \alpha_\te{d}'^{N^{(m)} - q },
    \end{align}
    where $ q \in \{ 0, 2, ..., 2\lfloor \frac{N^{(m)}}{2} \rfloor  \} $ and $N^{(m)}$ is the total number of photons in the $m$th virtual mode
    \begin{equation}
        N^{(m)} = \sum_{k=1}^K s_k^{(m)}.
    \end{equation}
    Here, $T_{k,m}$ is an element of the interferometer matrix \textbf{T}. It represents the amplitudes of the transformation  of a photon from the $m$th port to the $k$th port. Expressions for parameters $\beta_\te{d}'$ and $\alpha_\te{d}'$ are 
    \begin{align}
        & \alpha_\te{d} ' = \dfrac{ \eta_\mathrm{t}(1-\eta_\mathrm{ind})( 1 - \eta_\mathrm{t}(1-\eta_\mathrm{ind}) ) \sinh^2{r} }{ 1 + \eta_\mathrm{t}(1-\eta_\mathrm{ind}) (2 - \eta_\mathrm{t}( 1 - \eta_\mathrm{ind} )) \sinh^2{r}}, \label{alpha_d_prime}
        \\
        &\beta_\te{d}' = \dfrac{ \eta_\mathrm{t}(1-\eta_\mathrm{ind}) \sinh{r} \cosh{r} }{ 1 + \eta_\mathrm{t}(1-\eta_\mathrm{ind}) ( 2 - \eta_\mathrm{t}( 1 - \eta_\mathrm{ind} )) \sinh^2{r}}. \label{beta_d_prime}
    \end{align}
    A detailed derivation of Eq.(\ref{P_dis}) can be found in the Methods section . Hence the computational complexity of calculating one specific probability from $m$th virtual mode is only a 1st degree polynomial respect to total number of output photons. 
    
    It should be noted that this result does not indicate that the exact calculation of $P_{\mathrm{dis}} $ can be done in polynomial time, because it contains $\prod_{k=1}^K \frac{(M - 1 + s_k)!}{ (M-1)! s_k! }$ terms. Each term corresponds to a possible combination of $\mathbf{s}^{(1)},...,\mathbf{s}^{(M)}$. For the extreme case of the collision-free regime, there would still be $ M^N $ terms. Even though the computational complexity of calculating each term is only polynomial according to Eq.(\ref{P_dis}), adding them costs at least exponential time.
    Nevertheless, unlike Eq.(\ref{indisHaf}), Eq.(\ref{P_dis}) provides a back door for classical simulation. In the next section we propose an classical simulation method for generating $P_{\mathrm{dis}}$.
    
\subsection*{Efficient classical simulation for distinguishable GBS \label{sec:simulation}}    

     Looking at Eq.(\ref{P_dis}) closely, we find that by multiplying an additional factorial, the product on the right side forms a multinomial distribution:
    \begin{equation}
        P_\te{prod}^{(m)}(\mathbf{s}) = 
        N! \prod_{k=1}^K \dfrac{|T_{k,m}|^{2s_k}}{{s_k}!} \label{Pr_s},
    \end{equation}
    which means that the output port of each photon is randomly chosen among all $K$ ports following the probability distribution $( |T_{1,m}|^2, |T_{2,m}|^2 ... |T_{K,m}|^2 )$. $P_\te{prod}^{(m)}$(\textbf{s}) is also related to the probability distribution of obtaining output pattern \textbf{s} in distinguishable AABS \cite{aaronson_bosonsampling_2014}:
    \begin{equation}
        P_\te{AA}(\mathbf{s}) = \dfrac{\mathrm{Perm}(\mathbf{T}_\mathbf{s}^{\#})}{s_1! ... s_K!}\label{DisAABS},
    \end{equation}
    where $\mathbf{T}_\mathbf{s}^{\#}$ denotes a matrix with entries $|T_{ij}|^2$ where $T_{ij}$ is an entry of the original complex AABS transformation matrix $\mathbf{T_s}$. Under the condition that there is only one input port, Eq.(\ref{DisAABS}) reduces to Eq.(\ref{Pr_s}). With this in mind, the remaining coefficients in Eq.(\ref{P_dis}) can be interpreted as the probability of obtaining $N^{(m)}$ photons from the state described by Eq.(\ref{CVdis}):
    \begin{equation}
        P_{\mathrm{num}}^{(m)}(N) = \dfrac{ G(N) }{N!\sqrt{\det(\mathbf{Q}^{(m)})}}. \label{Pr_N}
    \end{equation}
    Now we can write Eq.(\ref{P_dis}) as a multiplication of Eq.(\ref{Pr_s}) and Eq.(\ref{Pr_N}).
    
    Such an analysis enables us to provide a classical sampling method for distinguishable GBS, with the help of the two probabilities distributions $P_\te{prod}^{(m)}$(\textbf{s}) and $P_\te{num}^{(m)}(N$). Before doing that, we need to set the range for the number of photons. Theoretically, this range is from zero to infinity, but since the probability decreases exponentially with $N$, and $G(N) \propto \alpha_\te{d}'^N$, it is convenient to truncate the sampling range at a number, $N_\te{t} = \overline{N}_\te{d} t$ where $\overline{N}_\te{d}$ is the average number of photons:
    \begin{equation}
        \overline{N}_\te{d} = \eta_\mathrm{t} (1-\eta_\mathrm{ind}) \sinh^2 r, 
    \end{equation}
    and $t$ is a truncation factor. Due to the truncation we renormalize $P_\te{num}^{(m)}(N)$ according to:
    \begin{equation}
        \widetilde{P}_\te{num}^{(m)}(N) = \dfrac{P_\te{num}^{(m)}(N)}{ \sum_{n=1}^{N_t} P_\te{num}^{(m)}(n) }.
    \end{equation}
    
%\begin{widetext}

\begin{algorithm}
\label{algo_Dis}
\SetKwData{Left}{left}
\SetKwData{This}{this}
\SetKwData{Up}{up}
\SetKwFunction{Union}{Union}
\SetKwFunction{FindCompress}{FindCompress}
\SetKwInOut{Input}{input}
\SetKwInOut{Output}{output}
\caption{Efficient sampling of distinguishable GBS}
    \Input{$M,K,N$, an $K\times K$ matrix $\mathbf{T}$ and $M$ \bc{probability mass functions (pmf)} $\widetilde{\mathbf{P}}_\te{num}^{(m)} = (\widetilde{P}_\te{num}^{(m)}(0),...,\widetilde{P}_\te{num}^{(m)}(N))$}
    \Output{a sample \textbf{s}}
    \BlankLine
    \textbf{s} $\leftarrow (s_1,...,s_K) $ \;
    $s_1 \leftarrow ... \leftarrow s_K \leftarrow 0$\;
    \For{ $ m \leftarrow 1 $ \KwTo $M$ }{
        $N^{(m)} \leftarrow$ Sample($\widetilde{\textbf{P}}_\te{num}^{(m)}$)
        \tcp*{ Sample a $N^{(m)}$ from ($0,1,...,N$) with pmf $\widetilde{\mathbf{P}}_\te{num}^{(m)}$}
        $\textbf{w}_m \leftarrow (|T_{1,m}|^2,...,|T_{K,m}|^2)$\;
        \For{$i \leftarrow 1 $ \KwTo $N^{(m)}$ }{
            $j \leftarrow $ Sample($\textbf{w}_m$) 
            \tcp*{ Sample a $j$ from ($1,...,K$) with pmf $\mathbf{w}_m$}
            $s_j \leftarrow s_j + 1$\;
        }
    } 
    \Return \textbf{s}
\end{algorithm}
    
%\end{widetext}   
    
    Algorithm 1 then allows us to sample the output pattern from all virtual modes. The computational complexity of the worst-case scenario is \bc{$O(MK \lceil t\overline{N}_d  \rceil)$} which scales only polynomially. We can create a probability distribution of all output patterns from distinguishable GBS with $\varepsilon$ accuracy requiring a computational cost \bc{$O( MK \lceil t\overline{N}_d \rceil / \varepsilon )$.} $n$ binary digit accuracy can be achieved for each probability if we let $\varepsilon = 1/2^n$. We denote this probability distribution as $P_{\mathrm{sim}}(\mathbf{s,\varepsilon})$.
    
    While $P_{\mathrm{dis}}(\textbf{s}) $ is only calculated for one \textbf{s} at a time, $P_{\mathrm{sim}}(\mathbf{s,\varepsilon})$ updates the probabilities for all output patterns simultaneously with each sampling. This is extremely useful in Eq.(\ref{regrouped_ProbPNR}) where $P_{\mathrm{dis}}(\textbf{s}) $ of a considerable number of different patterns \textbf{s} needs to be calculated to obtain the result. Naturally we obtain an approximation to $P(\textbf{s}) $ with accuracy $\varepsilon$ by replacing $P_{\mathrm{dis}}(\textbf{s}) $ with $P_{\mathrm{sim}}(\textbf{s},\varepsilon) $. 

    \rc{Algorithm 1 enables efficient sampling for the photons from the virtual modes, which makes it possible for us to reduce the computational time in the following probability approximation and simulation algorithms.}
    
\subsection*{Partially distinguishable GBS with threshold detectors \label{sec:threshold}}
    
    In the original proposal of GBS, each output port has a PNR detector. For a quantum supremacy demonstration, the number of ports --- and hence the number of PNR detectors --- is rather large, which may be prohibitive experimentally. As such, in works such as Ref. \cite{zhong_quantum_2020} they were replaced with threshold detectors where the detection result only shows the presence of the photons regardless of its exact number. If any number of photons greater or equal to one are detected, it is refered to as a “click". Of course, the probability of a click can be calculated by adding the probabilities of all possible output patterns from PNR detectors over an infinite number of terms. A better way than this direct approach is to use the P-functions of the POVM elements $|0\rangle\langle0|$ and $\hat{\textbf{I}} - |0\rangle\langle0|$, where one can directly calculate the probability of the output pattern \textbf{s} as in Ref. \cite{quesada_gaussian_2018}
    \begin{equation}
        P(\mathbf{s}) = \dfrac{ \mathrm{Tor}( \mathbf{Q}_{\mathcal{U}} )  }{\sqrt{\mathrm{det}(\mathbf{Q}) }}, \label{prob_Torontonian}
    \end{equation}
    with a matrix function defined as Torontonian:
    \begin{equation}
        \mathrm{Tor}( \mathbf{\mathbf{Q}}_{\mathcal{U}} ) =  \sum_{\mathcal{V} \in \mathcal{P}( \mathcal{U} ) } (-1)^{|\mathcal{U}| - |\mathcal{V}|} \dfrac{1}{ \sqrt{ \mathrm{det} \left( (\mathbf{Q}^{-1})_\mathcal{V} \right) } }. \label{Torontonian}
    \end{equation}
    Here, $\mathcal{U}$ is a set where the elements are the ports that have clicks in pattern \textbf{s}, $\mathcal{P}(\mathcal{U})$ is the power set which contains all the subsets of $\mathcal{U}$, and $\mathbf{Q}_{\mathcal{U}}^{(m)}$ is a matrix formed by keeping in $\mathbf{Q}^{(m)}$ only the rows and columns corresponding to the ports in set $\mathcal{U}$.
     
    In Ref. \cite{quesada_gaussian_2018}, Eq.(\ref{prob_Torontonian}) is obtained by considering only the Gaussian state of ideal GBS. Furthermore, how to include the effects of partial distinguishablility is not obvious from their formalism. Therefore, we need a new expression for $P(\mathbf{s})$ to include the effects of all $M$+1 modes and make Eq.(\ref{prob_Torontonian}) a special case. We propose the probability as a weighted sum of the marginal probabilities of no-click events:
    \begin{equation}
        P(\mathbf{s}) = \sum_{\mathcal{V} \in \mathcal{P}( \mathcal{U} ) } (-1)^{|\mathcal{U}| - |\mathcal{V}|}\  \widetilde{P}(\mathcal{R}), \label{shanghainian}
    \end{equation}
    where $\widetilde{P}(\mathcal{R})$ is the marginal probability of a no-click event for the ports in the given set $\mathcal{R}$ with the expression:
    \begin{equation}
        \widetilde{P}(\mathcal{R}) = \prod_{m'=0}^M\widetilde{P}^{(m')}(\mathcal{R}). \label{marginalProduct}
    \end{equation}
    Here, $\mathcal{R}$ is the set difference of $[1,M]$ and  $\mathcal{V}$ i.e. $\mathcal{R} = [1,M] - \mathcal{V} $. $\widetilde{P}^{(m')}(\mathcal{R})$ is the
      marginal probability of a no-click event in the $m'$th mode.

   For Gaussian states the marginal probability distribution of certain ports can be directly calculated by only considering the columns and rows corresponding to these ports in the covariance matrix. Additionally, the marginal probability of a no-click event is a function of determinant, therefore we have
    \begin{equation}
        \widetilde{P}^{(m')}(\mathcal{R}) = \dfrac{1}{\sqrt{\mathrm{det}\left( \mathbf{Q}_{\mathcal{R}}^{(m')} \right)}}. \label{margnialp}
    \end{equation}
    When $\eta_\mathrm{ind}=1$, all photons are indistinguishable hence there is no click in any $m$th virtual mode such that $\widetilde{P}^{(m)}(\mathcal{R})=1$. 
    By proving that 
    \begin{equation}
        \det(\mathbf{Q}_{\mathcal{R}}^{(0)}) = \det(\mathbf{Q}^{(0)}) \det\left( \left( { \mathbf{Q}^{(0)} }^{-1} \right)_\mathcal{V} \right), \label{equval_Torontonian}
    \end{equation}
    Eq.(\ref{prob_Torontonian}) becomes a special case of Eq.(\ref{shanghainian}). The proof is given in the Methods section. Apparently, $\det(\mathbf{Q}^{(0)})$ is irreducible so that the complexity of calculating $\widetilde{P}^{(0)}(\mathcal{R})$ remains $O(|\mathcal{R}|^3)$ as in the case of Torontonian.
    
    Interestingly, the marginal probability of a no-click events in all virtual modes can be reduced to
    \begin{equation}
        \widetilde{P}^{(m)}(\mathcal{R}) = \dfrac{1}{\sqrt{(1+ \mathcal{T}_m(\mathcal{R}) \alpha_\te{d} )^2 - (\mathcal{T}_m(\mathcal{R}) \beta_\te{d})^2 }}, \label{tor_dis}
    \end{equation}
    where $\mathcal{T}_m(\mathcal{R}) = \sum_{j\in\mathcal{R}}|T_{j,m}|^2 $. A detailed derivation of Eq.(\ref{tor_dis}) can be found in the Methods section. $\mathcal{T}_m(\mathcal{R})$ can be interpreted as the total transmission rate of ports in $\mathcal{R}$. When $\eta_\mathrm{ind}=1$, $\alpha_\te{d} = \beta_\te{d} = 0$ we have $ \widetilde{P}^{(m)}(\mathcal{R})=1$.
    This gives an exact calculation of the probability distribution of threshold detector GBS with partial distinguishability. 
   
\subsection*{ Quantum supremacy with partially distinguishable GBS}    
    
    We have obtained the probability distribution of partially distinguishable GBS that does not require any assumptions of being collision-free, or having a determinate photon number and explicitly include losses. We proceed to discuss how partial distinguishability affects quantum supremacy. 
    
    Let us define an approximate version of the probability $P(\textbf{s})$ by replacing $P_\te{dis}(\mathbf{s})$ with $P_{\mathrm{sim}}(\mathbf{s},\varepsilon)$ and truncating Eq.(\ref{regrouped_ProbPNR}) at a certain value $N_\te{cut}$
    \begin{align}
        &P_\te{approx}(\mathbf{s},\widetilde{\varepsilon},N_\te{cut}) = \sum_{n=0}^{N_\te{cut}} P_n(\mathbf{s},\varepsilon_n)\label{P_approx},
        \\
        &P_n(\mathbf{s},\varepsilon_n) = \sum_{\mathbf{s}_{n}^{(0)}} 
        P^{(0)}(\mathbf{s}_n^{(0)})
        P_{\mathrm{sim}}(\mathbf{s} - \mathbf{s}_n^{(0)},\varepsilon).\label{P_approx_part}
    \end{align}
    \bc{
    Here we can write the accuracy for  $P_\mathrm{approx}$ and $P_n$  respectively as
    \begin{align*}
         & \widetilde{\varepsilon} = \varepsilon \sum_{n=0}^{N_\te{cut}} \sum_{\mathbf{s}_{n}^{(0)}} 
        P^{(0)}(\mathbf{s}_n^{(0)})  \\
                & \varepsilon_n = \varepsilon \sum_{\mathbf{s}_{n}^{(0)}} 
        P^{(0)}(\mathbf{s}_n^{(0)})  . 
    \end{align*}
   We note that the accuracy here refers to the probability for a particular output pattern rather than the whole distribution. 
    Obviously  $\sum_{n=0}^{N_\te{cut}} \sum_{\mathbf{s}_{n}^{(0)}}        P^{(0)}(\mathbf{s}_n^{(0)}) $ occupies a tiny portion of the whole distribution. Therefore we have $\varepsilon_n \ll \varepsilon$ and $\widetilde{\varepsilon} \ll \varepsilon$. Hence the absolute accuracy of  $P_\mathrm{approx}$  is larger compared to that of $P_\mathrm{sim}$  although the relative accuracy most likely stays unchanged. 
    }

    Here, $P_n(\mathbf{s},\varepsilon_n)$ includes the contributions of all configurations that have $n$ indistinguishable photons. The magnitude of this term depends on the indistinguishability efficiency $\eta_\te{ind}$. For the extreme condition that $\eta_\mathrm{ind} = 1$ which corresponds to the ideal case, $P_n(\mathbf{s},\varepsilon_n)$ from $n < N$ are  all zero. Since the dependence of $P_n(\mathbf{s},\varepsilon_n)$ on $n$ for $\eta_\te{ind} < 1$ is approximately exponentially decreasing, we may safely truncate $P_\te{approx}$ with $N_\te{cut}$ smaller than $N$. \bc{The fidelity of the approximation is defined as }
    \begin{equation}
        F(\mathbf{s},\widetilde{\varepsilon}, N_\te{cut}) = P_\te{approx}(\mathbf{s},\widetilde{\varepsilon}, N_\te{cut}) / P(\mathbf{s}).
        \label{fidelitydef}
    \end{equation}
    This approximation is powerful because the computational complexity increases superexponentially with $N_\te{cut}$. \bc{This is because the computational complexity of calculating $P^{(0)}(\mathbf{s}_n^{(0)})$ is $O(n^3 2^n)$ which is exponential and the number of elements in $\mathbf{s}_{n}^{(0)}$ is maximally $\binom Nn$ when the output pattern is collision-free. The complexity of $P_\te{approx}(\mathbf{s},\widetilde{\varepsilon}, N_\te{cut})$ is then at most $O( N^{N_\mathrm{cut}} 2^{N_\te{cut}} )$ which is polynomial to $N$. By using the $N_\mathrm{cut}$ parameter, this reduces the computational cost of the approximation, reducing it from the ideal GBS case which takes $O(N^3 2^N)$ steps. From Fig. 3 we see that with $N_\mathrm{cut}=2$ and $N=9$, which has only a modest computational overhead,  the fidelity of our approximation can be maintained to exceed $98.2\%$. This demonstrates how effective our approximation method is.
    }

    \begin{figure}[t]
        \includegraphics[width=0.5\textwidth]{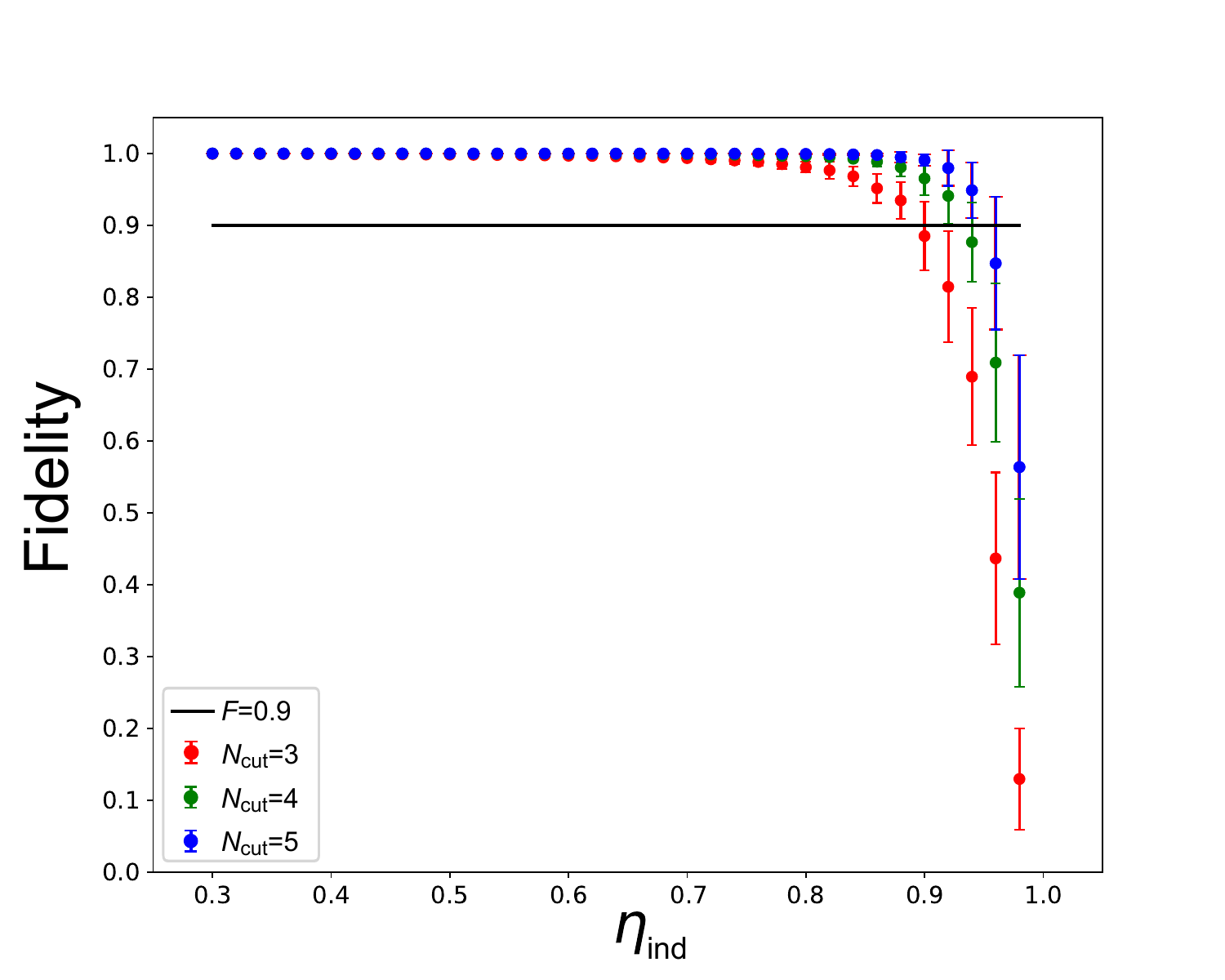}
        \caption{\label{fig:x_series} \raggedright Numerical evaluations of fidelity (\ref{fidelitydef}) for $N_\te{cut}$=3 , $N_\te{cut}$=4 and $ N_\te{cut}$=5  when $K$=35, $M$=6, $N$=6, $r$=0.9, $\eta_\mathrm{t}$=0.9. The output pattern is $s_1$=...=$s_N$=1, $s_{N+1}$ =...= $s_K$ = 0 the other choices of output pattern give similar results. This result is obtained over 94 Haar-random unitary matrices.  }
    \end{figure}
    
    We have not been able to obtain an analytical relationship between $F$ and $\eta_\mathrm{ind}$. However, numerically we observe that the fidelity obeys an approximate relation
    \begin{equation}
        F \approx 1-c\mathrm{e}^{\eta_\mathrm{ind}},
    \end{equation} 
    as shown in Fig. 2, where $c$ is a fitting parameter. According to this relation, the approximation can achieve high fidelity for $N_\te{cut}$ much smaller than $N$ unless $\eta_\mathrm{ind}$ is close to 1. In Fig. 2 we show the approximation for various $N_\te{cut}$. We see that the fidelity is above 0.9 even for a modest $N_\te{cut}=3$ approximation with $\eta_\te{ind}$ as large as 0.9. 

    \begin{figure*}[t]
    %\centering
    \begin{subfigure}[b]{0.49\textwidth}
        \includegraphics[width=\textwidth]{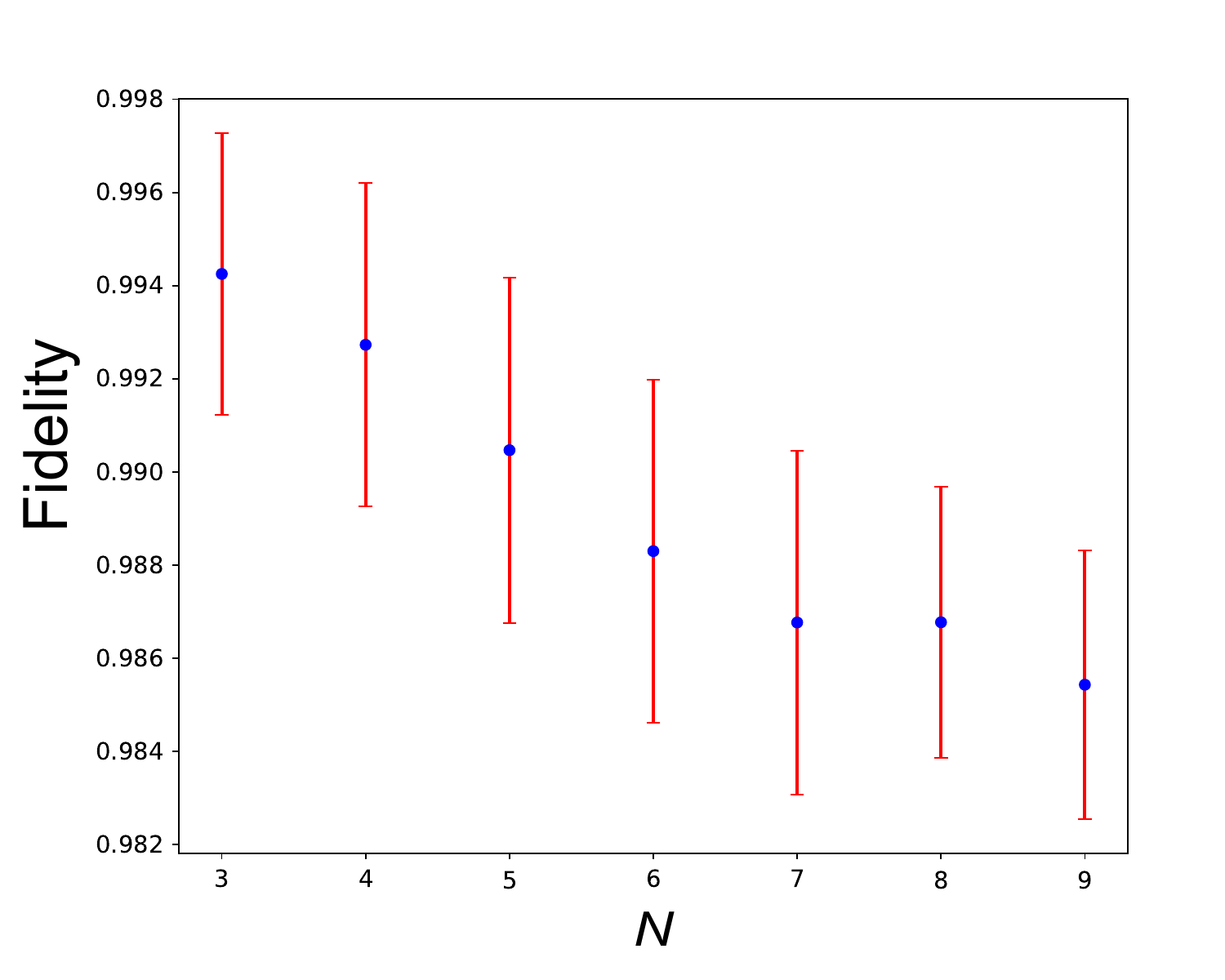}
        \caption{}
        \label{fig:n_series}
    \end{subfigure}
    \hfill
    \begin{subfigure}[b]{0.5\textwidth}
        \includegraphics[width=\textwidth]{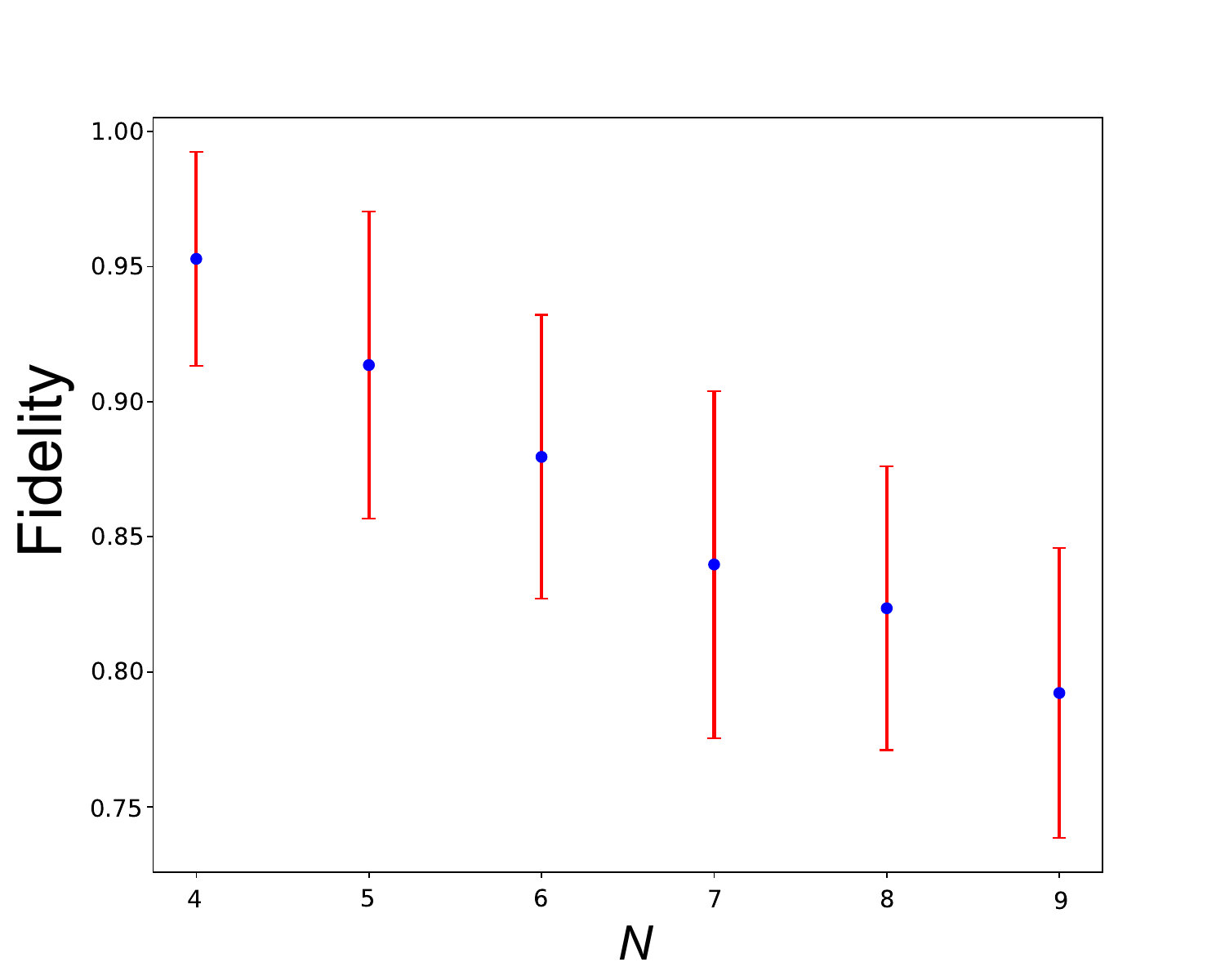}
        \caption{}
    \end{subfigure}
        % Here is how to import EPS art
        \caption{\raggedright Numerical evaluations of fidelity (\ref{fidelitydef}) for $P_\te{approx}$ with number of detected photons $N$ for (a) $N_\te{cut}=2$, $K=30$, $M=6$, $r$=0.9, $\eta_\mathrm{ind}$=0.5, $\eta_\mathrm{t}$=0.9; (b) $N_\te{cut}$=3, $K$=30, $M$=6, $r$=0.9, $\eta_\mathrm{ind}$=0.9, $\eta_\mathrm{t}$=0.99. The output pattern for each $N$ is $s_1=...=s_N=1, s_{N+1}=...=s_{K}=0$, the other choices of output pattern give similar results. These results are obtained over 94 Haar-random unitary matrices.  }
    \end{figure*}

    Another observation is that for a given GBS setup, when the total number of photons of an output pattern increases while the order of an approximation $N_\te{cut}$ is fixed, the fidelity of the $P_\te{approx}$ only decreases linearly or even more slowly. For the example shown in Fig. 3, we use $P_\te{approx} $ with $ N_\te{cut}=2$ to approximate the output patterns for a total number of photons $N$, from 3 to 9. While the exact calculation of these output patterns requires an exponential increase in computational time, the fidelity of the approximation $P_\te{approx}$ decreases linearly, instead of exponentially. The rate of decrease in fact appears to ease as $N$ increases. \bc{Performing a linear extrapolation of the data for $N<8$ to $ N = 50 $, we expect a fidelity of about $ 90 \% $, which is likely to be an underestimate as the data has some evidence of reducing in gradient with $N$. } This relation between fidelity and computational time again shows the effectiveness of this approximation method.

    For GBS with threshold detectors, partial distinguishability does not affect quantum supremacy directly because the main exponential contribution to the complexity comes from the number of elements to be calculated, which is $2^N$. Partial distinguishability only reduces the complexity of calculating one element from $O(N^3)$ to $O(N)$, by converting calculation of the determinant to multiplication as shown in Eq.(\ref{tor_dis}). 
   We note that if we let $\eta_\mathrm{ind}\approx 0$ such that all photons are almost distinguishable, the computational complexity is still exponential because the number of elements is still $2^N$ which is established as the proof for exponential complexity \bc{of calculating probability for a particular output in GBS with threshold detectors. But as we have seen above, the sampling method for $P_{\mathrm{sim}}(\mathbf{s},\varepsilon)$ can be used to efficiently sample the distribution in this limit. }
   
   \rc{We note that while we have focused primarily on the probability of an output pattern in this paper, our model can also be used for generating samples for partially distinguishable GBS. Our model are compatible with existing state-of-the-art simulation algorithms\cite{quesada_quadratic_2021,bulmer_boundary_2021}, which enable us to take advantages of these algorithms and the imperfect indistinguishability at the same time. For GBS with PNR detectors, we first create two samples. One is for photons from all $M$ distinguishable modes, directly created by Algorithm 1. The other is for the photons from  the indistinguishable mode, created by feeding the covariance matrix representing this mode, $\begin{bmatrix}
         \mathbf{T} & \mathbf{0}
         \\ \mathbf{0} & \mathbf{T}^*
        \end{bmatrix}
        \widetilde{\textbf{V}}^{(0)}
        \begin{bmatrix}
         \mathbf{T^\dag} & \mathbf{0}
         \\\mathbf{0} & \mathbf{T}^T
        \end{bmatrix}$, into the sampling method provided in Ref.\cite{quesada_quadratic_2021}.  
        Then, we combine these two samples by adding the results of each port to create one final sample.}
        
        \rc{For GBS with threshold detectors, naturally we can directly take the sample created for PNR detectors. Nevertheless, we can improve the sampling algorithm by taking advantages of the following two reasons. One reason is that for threshold detectors, one photon is enough to register a click in a port, whether this photon comes from the indistinguishable mode or distinguishable modes is irrelevant. The other reason is that the complexity of sampling an indistinguishable photon is much larger than that of sampling a distinguishable one. Therefore, we can first sample the $M$ distinguishable modes. According to that sampling result, then we can neglect the output ports where clicks are already registered. Hence we only need to sample the output pattern for the remaining ports. This is possible for Gaussian state because the marginal probabilities can be calculated by selecting the corresponding rows and columns of the covariance matrix. The details of the algorithm can be found in Algorithm 2. } 
   
   \begin{algorithm}
\label{algo_Dis_threshold}
\SetKwData{Left}{left}
\SetKwData{This}{this}
\SetKwData{Up}{up}
\SetKwFunction{Union}{Union}
\SetKwFunction{FindCompress}{FindCompress}
\SetKwInOut{Input}{input}
\SetKwInOut{Output}{output}
\caption{Sampling algorithm of partially distinguishable GBS with threshold detectors}
    \Input{$M,K,N$, a set $\mathcal{R}=\{1,2,...,K\}$, an $K\times K$ transformation matrix $\mathbf{T}$, an $2K \times 2K$ covariance matrix of the input light from indistinguishable mode $\widetilde{V}^{(0)}$ and $M$ \bc{probability mass functions (pmf)} $\widetilde{\mathbf{P}}_\te{num}^{(m)} = (\widetilde{P}_\te{num}^{(m)}(0),...,\widetilde{P}_\te{num}^{(m)}(N))$,}
    \Output{a sample \textbf{s}}
    \BlankLine
    create a sample \textbf{s} = ($s_1,...,s_K$) by inputting $M,K,N,\mathbf{T},\widetilde{\mathbf{P}}_\te{num}^{(m)}$ into \textbf{Algorithm 1}\;
    \For{$ k \leftarrow 1 $ \KwTo $K$}{\If{$s_k>0$}{$s_k \leftarrow 1$\; $\mathcal{R}\leftarrow\mathcal{R}-\{k\}$\; }}
    calculate $\begin{bmatrix}
         \mathbf{T} & \mathbf{0}
         \\ \mathbf{0} & \mathbf{T}^*
        \end{bmatrix}
        \widetilde{\textbf{V}}^{(0)}
        \begin{bmatrix}
         \mathbf{T^\dag} & \mathbf{0}
         \\\mathbf{0} & \mathbf{T}^T
        \end{bmatrix}$, denoted as \textbf{V}\;
    create a selected matrix $\mathbf{V}_\mathcal{R}$ by keeping the rows and columns of \textbf{V} according to the ports listed in $\mathcal{R}$\;
    create a sample \textbf{t} = $(t_1,...,t_{|\mathcal{R}|})$ only for the ports listed in $\mathcal{R}$ by inputting $\mathbf{V}_\mathcal{R}, K, N$ into the sampling algorithm in Ref.\cite{bulmer_boundary_2021}\;
    \For{$i\leftarrow 1$ \KwTo $|\mathcal{R}|$}{\If{$t_i>0$}{$j \leftarrow i$th element of $\mathcal{R}$\; $s_j \leftarrow 1$\; }}
    \Return \textbf{s}
\end{algorithm}

     \rc{The complexity of generating a sample with $N$ photons(clicks) for ideal GBS are $O(N^32^{N/2})$ in the two state-of-the-art algorithms we used.  For a partially distinguishable case, since only part of the sampled photons actually comes from the indistinguishable mode, the complexity of will be reduced. For every photon counted as a photon in PNR detectors or registered as a click in threshold detectors, approximately $\eta_\mathrm{ind}$ of the probability it comes from the indistinguishable mode, which makes the complexity $O((N\eta_{\mathrm{ind}})^32^{N\eta_{\mathrm{ind}}/2})$. This expression covers quantitatively all different indistinguishability from to totally distinguishable case where $\eta_\mathrm{ind} = 0$ that the exponential term becomes one, to the perfectly indistinguishable case where $\eta_\mathrm{ind}=1$ that the complexity remains the same as original. For partially distinguishable cases, There is a exponential reduction of $2^{N(1-\eta_\mathrm{ind})/2}$. This reduction of complexity can be substantial with large number of sampled photons(clicks) or comparatively small indistinguishable efficiency. It indicates that $1/\eta_\mathrm{ind}$ as many photons(clicks) are needed in a partially distinguishable GBS experiment to reach the regime where classical simulations become intractable. } 
    
\section*{Discussion}
    
    In this paper we have developed a model which allows us to model GBS with partially distinguishable photons and obtain the expressions for the probabilities of a given output pattern for both PNR and threshold detectors. The model is based on the construction of virtual modes which incorporates the distinguishable photons and forms a new Gaussian state that propagates inside the interferometer independently until it reaches the detectors. We have proved that the expression for the probability of the photons from these distinguishable Gaussian states contains the previous result obtained by Aaronson and Arkhipov for the distinguishable AABS as a special case. This is because we included the indeterminate-photon-number nature of Gaussian states in contrast to AABS where the photon number is fixed. Based on that we proposed an algorithm for efficient simulation of the output patterns from these distinguishable Gaussian states which enables us to exponentially reduce the computational time of calculating the probabilities.
    
    Our method provides a framework to calculate the probabilities for imperfect GBS, especially for GBS with threshold detectors which has only been theoretically investigated for the ideal case. We proved that the Torontonian --- the result obtained in the ideal case --- is a special case within our framework. We note that for low indistinguishability, the proof that supports the complexity of computing the Torontonian still holds. Our aforementioned simulation algorithm can 
   reduce the computationally hard exact calculation with a highly accurate approximation particularly for low indistinguishability.
    
    For GBS with PNR detectors, which to date is more theoretically developed, we proposed an approximation based on the structure of the expression of the probability with respect to indistinguishability efficiency to show how partial distinguishability affects quantum supremacy. We have taken advantage of the physical nature of the indistinguishability, i.e. interference of photons which is the cause of the computationally-hard complexity. Therefore for low indistinguishability, we only include contributions from a smaller number of interfering photons which takes exponentially less time. \bc{We note that for GBS with extremely high indistinguishability a direct calculation of the Hafnian is more efficient than the approximation method due to the additional overhead of incorporating  the distinguishable photons.}   Numerically we showed how the computational time of our approximation for a given fidelity decreases exponentially with reduced indistinguishablity which indicates the relationship between partial distinguishablity and quantum supremacy.

    While we have obtained the expression Eq.(\ref{shanghainian}) for the exact calculation of the probability for GBS with threshold detectors, using this to find an approximation method is less easily constructed, unlike the case for PNR detectors. For the PNR detector case, the approximation Eq.(\ref{P_approx}) comes naturally from the expression Eq.(\ref{ProbPNRtotal}) for exact sampling. In this sense, GBS with threshould detectors is advantageous than GBS with PNR detectors in the context of probability calculation due to the lack of an efficient approximation algorithm. Though
    For PNR detectors, our approximation method may be used to obtain the bound for indistinguishability efficiency for efficient simulation at a required fidelity and computational time. In our numerical studies we have found that the indistinguishability efficiency should be larger than 0.9, such that the computational requirements of the approximation method become costly. This is satisfied for the recent GBS experiments to date \cite{zhong_experimental_2019,paesani_generation_2019,zhong_quantum_2020,zhong_phase-programmable_2021}. 
    
    \rc{Our model also provides a foundation for taking advantage of the partial distinguishability to reduce the computational time in classical simulation of GBS. We propose two simulation algorithms for partially distinguishable GBS with both types of detectors based on Algorithm 1 and corresponding state-of-the-art algorithm. With imperfect indistinguishability, the computational time will exponentially reduced by depending on the number of photons(clicks) and how imperfect indistinguishability is. Our result shows that roughly $1/\eta_\mathrm{ind}$ as many as photons(clicks) are needed in a experiment where compared to a classical simulation. Therefore, in both probability calculation and classical simulation, our model shows that partial distinguishability can affect quantum supremacy in GBS.}

\section*{Methods}
    
\subsection*{Q-function covariance matrices and kernel matrices of all $M$+1 modes \label{App:kernel}} 
    \bc{
    As shown in Eq.(\ref{oriHaf}), the probability of obtaining a certain output pattern \textbf{s} from a Gaussian state requires the covariance matrix of the Q-function of the state, denoted as $\mathbf{Q}$. It is needed for the value of its determinant and for constructing the kernel matrix \textbf{A}:
    \begin{equation}
        \mathbf{A} = 
        \begin{bmatrix}
            \mathbf{0} & \mathbf{I}_K
            \\
            \mathbf{I}_K & \mathbf{0}
        \end{bmatrix}
        \left(
            \mathbf{I}_{2K} - \mathbf{Q}^{-1}
        \right) , 
        \label{kernelDefine}
    \end{equation}
    where $\mathbf{Q}$ is converted from the real covariance matrix \textbf{V}.
    From the relation between the covariance matrix of the light before entering the interferometer, namely $\textbf{Q}_{\text{in}}$, and the light after the interferometer, namely $\textbf{Q}_{\text{out}}$:
    \begin{equation}
        \textbf{Q}_{\text{out}} = \begin{bmatrix}
         \mathbf{T} & \mathbf{0}
         \\ \mathbf{0} & \mathbf{T}^*
        \end{bmatrix}
        \textbf{Q}_{\text{in}}
        \begin{bmatrix}
         \mathbf{T^\dag} & \mathbf{0}
         \\\mathbf{0} & \mathbf{T}^T
        \end{bmatrix},\label{trasformQ}
    \end{equation}
    we could obtain similar relation for the kernel matrices:
    \begin{equation}
        \mathbf{A}_{\text{out}} = 
        \matrixBlock{\mathbf{T}^*}{\mathbf{0}}{\mathbf{0}}{\mathbf{T}}
        \textbf{A}_{\text{in}}
        \matrixBlock{\mathbf{T}^\dag}{\mathbf{0}}{\mathbf{0}}{\mathbf{T}^T}.\label{traskernel}
    \end{equation}
    Eq.(\ref{traskernel}) is very useful because it allows us to calculate the kernel matrices successively from one interferometer to the other without the need of calculating the inverse matrix for every covariance matrix being transformed. The only kernel matrix we need to directly calculate from the definition Eq.(\ref{kernelDefine}) is for the light emerging from the source, which is what are we will calculate in the next step for our partially distinguishable light.}

    First we obtain $\textbf{Q}_{\text{in}}$ for all $M+1$ modes from Eq.(\ref{CVindis}) and Eq.(\ref{CVdis}):
    \begin{align}
        &\textbf{Q}_{\text{in}}^{(0)} = \textbf{I}_{2K} + \matrixBlock{\alpha_\mathrm{i}\textbf{J}^{(0)}}{\beta_\mathrm{i}\textbf{J}^{(0)}}{\beta_\mathrm{i}\textbf{J}^{(0)}}{\alpha_\mathrm{i}\textbf{J}^{(0)}},
        \\
        & \textbf{Q}_{\text{in}}^{(m)} = \textbf{I}_{2K} + \matrixBlock{\alpha_\mathrm{d}\textbf{J}^{(m)}}{\beta_\mathrm{d}\textbf{J}^{(m)}}{\beta_\mathrm{d}\textbf{J}^{(m)}}{\alpha_\mathrm{d}\textbf{J}^{(m)}},\label{disQin}
    \end{align}
    where 
    \begin{align}
        &\textbf{J}^{(0)} = \overbrace{\bigoplus 1 ... \bigoplus 1}^{M}\ \overbrace{\bigoplus 0 ... \bigoplus 0}^{K-M},
        \\
        &\textbf{J}^{(m)} = \overbrace{\bigoplus 0 ... \bigoplus 0}^{m-1}  \ \bigoplus 1\ \overbrace{\bigoplus 0 ... \bigoplus 0}^{K-m},
        \\
        &\alpha_\te{i} = \eta_\mathrm{t} \eta_\mathrm{ind} \sinh{r} \sinh{r},
        \\
        &\beta_\te{i} = \eta_\mathrm{t} \eta_\mathrm{ind} \sinh{r} \cosh{r},
        \\
        &\alpha_\te{d} = \eta_\mathrm{t}(1-\eta_\mathrm{ind}) \sinh{r} \sinh{r},\label{alpha_d}
        \\
        &\beta_\te{d} = \eta_\mathrm{t}(1-\eta_\mathrm{ind}) \sinh{r} \cosh{r}. \label{beta_d}
    \end{align}
    Then we calculate $\textbf{A}_{\mathrm{in}}$. By observing that 
    \begin{align}
        &\left( \mathbf{I}_{2K} + \matrixBlock{ \bigoplus_{j=1}^K \alpha_j }{ \bigoplus_{j=1}^K \beta_j }{ \bigoplus_{j=1}^K \alpha_j }{ \bigoplus_{j=1}^K \beta_j } \right) \nonumber
        \\
        &\cdot\left( \mathbf{I}_{2K} - \matrixBlock{ \bigoplus_{j=1}^K \alpha_j' }{ \bigoplus_{j=1}^K \beta_j' }{ \bigoplus_{j=1}^K \beta_j' }{ \bigoplus_{j=1}^K \alpha_j' } \right) = \mathbf{I}_{2K}, 
        \\
        \Rightarrow &
        \begin{cases}
            &\alpha_j' = 1 - \dfrac{1 + \alpha_j}{(1 + \alpha_j)^2 - \beta_j^2},
            \\
            &\beta_j' = \dfrac{\beta_j}{(1 + \alpha_j)^2 - \beta_j^2},\label{alphaprime}
        \end{cases}
    \end{align}
    it can be found that 
    \begin{align}
        &\textbf{A}_{\text{in}}^{(0)} =  \matrixBlock{\beta_\mathrm{i}'\textbf{J}^{(0)}}{\alpha_\mathrm{i}'\textbf{J}^{(0)}}{\alpha_\mathrm{i}'\textbf{J}^{(0)}}{\beta_\mathrm{i}'\textbf{J}^{(0)}},
        \\
        & \textbf{A}_{\text{in}}^{(m)} =  \matrixBlock{\beta_\mathrm{d}'\textbf{J}^{(m)}}{\alpha_\mathrm{d}'\textbf{J}^{(m)}}{\alpha_\mathrm{d}'\textbf{J}^{(m)}}{\beta_\mathrm{d}'\textbf{J}^{(m)}},
    \end{align}
    where the values of $\alpha_i', \alpha_d', \beta_i', \beta_d'$ can be calculated through Eqs.(\ref{alphaprime}):
    \begin{align}
        & \alpha_\mathrm{i}' = \dfrac{  (1-\eta_\mathrm{ind}\eta_\mathrm{t}) \eta_\mathrm{ind}\eta_\mathrm{t}\sinh^2{r}  }{ 1 + \eta_\mathrm{t}\eta_\mathrm{ind} (2 - \eta_\mathrm{t}\eta_\mathrm{ind}) \sinh^2{r}},
        \\
        &\beta_\mathrm{i}' = \dfrac{ \eta_\mathrm{t} \eta_\mathrm{ind} \sinh{r} \cosh{r} }{ 1 + \eta_\mathrm{t}\eta_\mathrm{ind} (2 - \eta_\mathrm{t}\eta_\mathrm{ind}) \sinh^2{r}},
        \\
        & \alpha_\te{d} ' = \dfrac{ \eta_\mathrm{t}(1-\eta_\mathrm{ind})( 1 - \eta_\mathrm{t}(1-\eta_\mathrm{ind}) ) \sinh^2{r} }{ 1 + \eta_\mathrm{t}(1-\eta_\mathrm{ind}) (2 - \eta_\mathrm{t}( 1 - \eta_\mathrm{ind} )) \sinh^2{r}},
        \\
        &\beta_\te{d}' = \dfrac{ \eta_\mathrm{t}(1-\eta_\mathrm{ind}) \sinh{r} \cosh{r} }{ 1 + \eta_\mathrm{t}(1-\eta_\mathrm{ind}) ( 2 - \eta_\mathrm{t}( 1 - \eta_\mathrm{ind} )) \sinh^2{r}}.
    \end{align}
In  the last step, we used Eq.(\ref{traskernel}) to obtain the kernel matrices after the propagation inside the interferometer:
    \begin{align}
        &\textbf{A}_{\text{out}}^{(0)} = \begin{bmatrix}
         \mathbf{T}^* & \mathbf{0}
         \\\mathbf{0} & \mathbf{T}
        \end{bmatrix}
        \matrixBlock{\beta_\mathrm{i}'\textbf{J}^{(0)}}{\alpha_\mathrm{i}'\textbf{J}^{(0)}}{\alpha_\mathrm{i}'\textbf{J}^{(0)}}{\beta_\mathrm{i}'\textbf{J}^{(0)}}
        \begin{bmatrix}
         \mathbf{T^\dag} & \mathbf{0}
         \\\mathbf{0} & \mathbf{T}^T
        \end{bmatrix},
        \\
        & \textbf{A}_{\text{out}}^{(m)} =  \begin{bmatrix}
         \beta_\te{d}' \mathbf{E}_{1,m}^* & \alpha_\te{d}'\mathbf{E}_{2,m}^*
         \\ \alpha_\te{d}'\mathbf{E}_{2,m} & \beta_\te{d}' \mathbf{E}_{1,m}
        \end{bmatrix}, \label{kernel_m}
    \end{align}
    where
    \begin{align}
        & \mathbf{E}_{1,m} = 
        \begin{bmatrix}
            T_{1,m}
            \\ \vdots
            \\ T_{K,m}
        \end{bmatrix}
        \begin{bmatrix}
            T_{1,m} & \cdots &T_{K,m}
        \end{bmatrix}, \label{E1m}
        \\
        & \mathbf{E}_{2,m} = 
        \begin{bmatrix}
            T_{1,m}
            \\ \vdots
            \\ T_{K,m}
        \end{bmatrix}
        \begin{bmatrix}
            T_{1,m}^* & \cdots &T_{K,m}^* 
        \end{bmatrix}. \label{E2m}
    \end{align}
    We note that the subscripts ``in'' and ``out'' are only used in this section because in the main text we are only interested in the kernel matrices after the interferometer.

\subsection*{Derivation of Eq.(\ref{P_dis}) \label{App:P_dis}}

    To obtain Eq.(\ref{P_dis}), firstly we need to calculate $\mathrm{Haf}(\mathbf{A}^{(m)})$ where the form of $\mathbf{A}^{(m)}$ is given by Eq.(\ref{kernel_m}). Direct calculation is very difficult so we observe that $\mathbf{A}^{(m)}$ can be decomposed into two matrices with no overlap:
    \begin{equation}
        \mathbf{A}^{(m)} = \beta_\te{d}'\matrixBlock{\mathbf{E}_{1,m}^*}{0}{0}{\mathbf{E}_{1,m}} + \alpha_\te{d}'\matrixBlock{0}{\mathbf{E}_{2,m}^*}{\mathbf{E}_{2,m}}{0}.
    \end{equation}
    Combined with the definition of Hafnian function we obtain
    \begin{equation}
        \mathrm{Haf}(\mathbf{A}^{(m)}) = \dfrac{1}{n!2^n} \sum_{\rho\in \mathrm{S}_{2n}} \prod_{j=1}^n G_{\rho(2j-1),\rho(2j)} H_{\rho(2j-1),\rho(2j)},
    \end{equation}
     where we have
     \begin{equation}
         \mathbf{G} = \matrixBlock{\mathbf{E}_{1,m}^*}{\mathbf{E}_{2,m}^*}{\mathbf{E}_{2,m}}{\mathbf{E}_{1,m}}
     \end{equation}
     and \textbf{H} is a $2n\times 2n$ matrix such that 
     \begin{equation}
         H_{i,j} = 
         \begin{cases}
            &\beta_\te{d}',\ \te{if}\ (i>n\ \te{AND} \ j>n)\ \te{OR} \ (i\le n\ \te{AND} \ j\le n)  
            \\
            &\alpha_\te{d}',\ \te{if}\ (i>n\ \te{AND}\ j\le n)\ \te{OR} \ (i\le n\ \te{AND}\ j> n)  .
        \end{cases}
     \end{equation}
    With the definitions Eq.(\ref{E1m}) and Eq.(\ref{E2m}), it is easy to prove that 
    \begin{equation}
        \prod_{j=1}^n G_{\rho(2j-1),\rho(2j)} = \prod_{i=1}^K |T_{k,m}|^{2s_k^{(m)}},
    \end{equation}
    which can be used to reduce the calculation of Haf($\mathbf{A}^{(m)}$) to the calculation of Haf(\textbf{H}):
    \begin{equation}
        \mathrm{Haf}(\mathbf{A}^{(m)}) = \prod_{k=1}^K |T_{k,m}|^{2s_k^{(m)}} \te{Haf}(\textbf{H}).
    \end{equation}
    
    The calculation of Haf(\textbf{H}) is non-trivial and we must resort to graph theory. As is well-known, the Hafnian is closely linked to weighted perfect matchings of a graph by the following definition for a $2n\times 2n$ matrix:
    \begin{equation}
        \te{Haf}(\mathbf{H}) = \sum_{\tau \in \te{PMP}(2n)} \prod_{(i,j)\in \tau} H_{i,j},
    \end{equation}
    which means we have to perfectly match $2n$ vertices to form one permutation in PMP($2n$). 
    
    Regarding the definition of $H_{i,j}$, we can divide these $2n$ vertices into two sets: $\mathcal{N}_1 = $ [$1,n$] and  $\mathcal{N}_1 = [n+1,2n] $. A match inside $\mathcal{N}_1$ or $\mathcal{N}_2$ corresponds to the weight $\beta_\te{d}'$ and the match between  $\mathcal{N}_1$ and $\mathcal{N}_2$ corresponds to the weight $\alpha_\te{d}'$. For a given permutation, if we denote the number of matches inside $\mathcal{N}_1$ or $\mathcal{N}_2$ as $q$, the number of matches between  $\mathcal{N}_1$ and $\mathcal{N}_2$ is consequently $n-q$. Note the value of $q$ is always even because a match in $\mathcal{N}_1$ inevitably leads to a match in $\mathcal{N}_2$. Now we can rewrite Haf(\textbf{H}) as a summation respect to $q$:
    \begin{equation}
        \te{Haf}(\mathbf{H}) = \sum_q f_q \beta_\te{d}'^q\alpha_\te{d}'^{n-q} ,  
    \end{equation}
    where $q \in \{ 0, 2, ..., 2\lfloor \frac{n}{2} \rfloor  \}$.
    
    The final step is to obtain the expression of $f_q$. 
    Firstly,  we regroup the vertices in $\mathcal{N}_1$ into two sets: $\mathcal{N}_{1,\te{in}}$ contains the vertices matching vertices inside $\mathcal{N}_1$; $\mathcal{N}_{1,\te{out}}$ contains the vertices matching vertices in $\mathcal{N}_2$. We do the same thing for vertices in $\mathcal{N}_2$ such that in total there are $\left(\dfrac{n!}{q!(n-q)!}\right)^2$ configurations for this process.
    
    Secondly, we count the number of perfect matches for vertices in $\mathcal{N}_{1,\te{in}}$ and $\mathcal{N}_{2,\te{in}}$. Since each one contributes $|$ PMP($q$) $|$ = $ (q-1)!! $ permutations, in total there are $ ((q-1)!!)^2 $ configurations in this process.
    
    Thirdly, we count the number of matches between vertices in $\mathcal{N}_{1,\te{out}}$ and $\mathcal{N}_{2,\te{out}}$. Since there are $n-q$ vertices in each set, we can straightforwardly obtain the number of configurations to be $(n-q)!$.
    
    Combining them we obtain the closed-form expression of $f_q$:
    \begin{equation}
        \begin{split}
            f_q &= \left(\dfrac{n!}{q!(n-q)!}\right)^2 ((q-1)!!)^2 (n-q)! 
            \\
            &= \dfrac{(n!)^2}{(q!!)^2(n-q)!}.
        \end{split}
    \end{equation}
    It is easy to prove that $\sum_q f_q = (2n - 1)!!$, which verifies the correctness of the expression. 
    
    Thus we obtain the analytical result for Haf($\mathbf{A}^{(m)}$) and hence Eq.(\ref{P_dis}) where we let $G(N)=$ Haf(\textbf{H}). 

\subsection*{ Proof of Eq.(\ref{equval_Torontonian}) \label{App:Torontonian}}
        This section proves that for a covariance matrix $\Sigma$ of size $2K\times2K$, $\mathcal{U} = [1,K]$, $\mathcal{R}$ is an arbitrary subset of $\mathcal{U}$, $\mathcal{R}^c$ is the relative complement set of $\mathcal{R}$ respect to $\mathcal{U}$, we will always have:
        \begin{equation}
            \det( \Sigma_\mathcal{R} ) = \det( \Sigma ) \det\left( (\Sigma^{(-1)})_{\mathcal{R}^c} \right). \label{torapp0}
        \end{equation}
        
        \textit{Proof}: First, we regroup the matrix $\Sigma$ as $\begin{bmatrix} \textbf{A} & \textbf{B} \\ \textbf{C} & \textbf{D} \end{bmatrix}$ by moving rows and columns so that \textbf{A} includes the indices from $\mathcal{R}^c $ and \textbf{B} includes the indices from $\mathcal{R}$. Note the sign of the determinant will not be changed because the interchange of rows (columns) are always carried out an even number of times therefore we have
        \begin{equation}
            \det(\Sigma) = \det( \matrixBlock{\textbf{A}}{\textbf{B}}{\textbf{C}}{\textbf{D}} ) . 
        \end{equation}
        
        Next, we use the Schur complement 
        \begin{equation}
            \det( \matrixBlock{\textbf{A}}{\textbf{B}}{\textbf{C}}{\textbf{D}} ) 
            = \det( \textbf{A} - \textbf{B}\textbf{D}^{-1}\textbf{C} ) \det(\textbf{D}),
        \end{equation}
        to obtain 
        \begin{equation}
            \det( \textbf{A} - \textbf{B}\textbf{D}^{-1}\textbf{C} ) 
            = \dfrac{\det(\Sigma)}{\det(\Sigma_\mathcal{R})}. \label{torapp1}
        \end{equation}
        
        Then we calculate the inverse matrix of $\matrixBlock{\textbf{A}}{\textbf{B}}{\textbf{C}}{\textbf{D}}$ to be $\matrixBlock{ (\textbf{A} - \textbf{B}\textbf{D}^{-1}\textbf{C})^{-1} }
            { -\textbf{A}^{-1}\textbf{B} (\textbf{A} - \textbf{B}\textbf{D}^{-1}\textbf{C})^{-1} }
            { - \textbf{D}^{-1}\textbf{C}(\textbf{D} - \textbf{C}\textbf{A}^{-1}\textbf{B})^{-1}  }
            {(\textbf{D} - \textbf{C}\textbf{A}^{-1}\textbf{B})^{-1}}. $
        Therefore
        \begin{equation}
            (\textbf{A} - \textbf{B}\textbf{D}^{-1}\textbf{C})^{-1}
            =\left( \Sigma^{-1} \right)_{\mathcal{R}^c},
        \end{equation}
        such that we obtain another equation of $\det( \textbf{A} - \textbf{B}\textbf{D}^{-1}\textbf{C} ) $
        \begin{equation}
            \det( \textbf{A} - \textbf{B}\textbf{D}^{-1}\textbf{C} ) 
            = \dfrac{1}{ \det\left( \left( \Sigma^{-1} \right)_{\mathcal{R}^c} \right) }. \label{torapp2}
        \end{equation}
        
        Combining Eq.(\ref{torapp1}) and Eq.(\ref{torapp2}) we obtain Eq.(\ref{torapp0}).

\subsection*{ Calculation of marginal probability of distinguishable photons \label{App:marginal_dis} }
    This section calculates the marginal probability of given ports according to Eq.(\ref{margnialp}). Basically, that amounts to calculating the determinant of a selected covariance matrix $\mathbf{Q}_\mathcal{R}^{(m)}$ whose rows and columns are selected from $\mathbf{Q}^{(m)}$ according to the ports listed in set $\mathcal{R}$. Without loss of generality, we presume there are $n$ elements in $\mathcal{R}$ and we denote its $i$th element as $R_i$. 
    
    \bc{First we need to calculate $\textbf{Q}^{(m)}$ which is $\textbf{Q}_{\mathrm{out}}^{(m)}$ obtained by putting Eq.(\ref{disQin}) into Eq.(\ref{trasformQ}):
    \begin{align}
        \mathbf{Q}_\mathrm{out}^{(m)} = \mathbf{I}_{2K} + 
        \begin{bmatrix}
         \alpha_\te{d} \mathbf{E}_{2,m} & \beta_\te{d} \mathbf{E}_{1,m}
         \\
         \beta_\te{d} \mathbf{E}_{1,m}^* & \alpha_\te{d} \mathbf{E}_{2,m}^*
        \end{bmatrix},\label{sigma_dis}
    \end{align}
    }
    \begin{widetext}
        Therefore, $\mathbf{Q}_{\mathcal{R}}^{(m)}$ can be written as a single matrix:
    \begin{equation}
        \mathbf{Q}_{\mathcal{R}}^{(m)} =
        \begin{bmatrix}
         (\gamma_1+\alpha_\te{d})T_{R_1,m}T_{R_1,m}^* & ... &\alpha_\te{d} T_{R_1,m}T_{R_n,m}^* & \beta_\te{d} T_{R_1,m}T_{R_1,m} & ... & \beta_\te{d} T_{R_1,m}T_{R_n,m}
         \\
         \vdots & \ddots & \vdots & \vdots & \ddots & \vdots
         \\ \alpha_\te{d} T_{R_n,m}T_{R_1,m}^* & ... & (\gamma_n+\alpha_\te{d})T_{R_n,m}T_{R_n,m}^* & \beta_\te{d} T_{R_n,m}T_{R_1,m} & ... & \beta_\te{d} T_{R_n,m}T_{R_n,m}
         \\
         \beta_\te{d} T_{R_1,m}^* T_{R_1,m}^* & ... & \beta_\te{d} T_{R_1,m}^*T_{R_n,m}^* & (\gamma_1+\alpha_\te{d}) T_{R_1,m}^*T_{R_1,m} & ... & \alpha_\te{d} T_{R_1,m}^* T_{R_n,m}
         \\
         \vdots & \ddots & \vdots & \vdots & \ddots & \vdots
         \\
         \beta_\te{d} T_{R_n,m}^* T_{R_1,m}^* & ... & \beta_\te{d} T_{R_n,m}^*T_{R_n,m}^* & \alpha_\te{d} T_{R_n,m}^*T_{R_1,m} & ... & (\gamma_K+\alpha_\te{d}) T_{R_n,m}^* T_{R_n,m},
        \end{bmatrix}
    \end{equation}
    where $\gamma_i = 1/|T_{R_i,m}|^2$. Now we take all the coefficients and form another matrix $\mathbf{C}$:
    \begin{equation}
        \mathbf{C} =
        \begin{bmatrix}
         \gamma_1+\alpha_\te{d} & ... &\alpha_\te{d}  & \beta_\te{d} & ... & \beta_\te{d} 
         \\
         \vdots & \ddots & \vdots & \vdots & \ddots & \vdots
         \\ \alpha_\te{d}  & ... & \gamma_K+\alpha_\te{d} & \beta_\te{d}  & ... & \beta_\te{d} 
         \\
         \beta_\te{d}  & ... & \beta_\te{d}  & \gamma_1+\alpha_\te{d}  & ... & \alpha_\te{d} 
         \\
         \vdots & \ddots & \vdots & \vdots & \ddots & \vdots
         \\
         \beta_\te{d}  & ... & \beta_\te{d}  & \alpha_\te{d}  & ... & \gamma_K+\alpha_\te{d}
        \end{bmatrix}.
    \end{equation}
    
    \end{widetext}
    
    Then we calculate its determinant by resorting to the definition of determinant
    \begin{equation}
        \begin{split}
            \mathrm{det}(\mathbf{Q}_\mathcal{R}^{(m)}) = &\sum_{\rho\in\mathrm{S}_{2K}} \te{Sgn}(\rho) \prod_{i=1}^{2n}\mathbf{Q}_\mathcal{R}^{(m)}(i,\rho(i))
            \\
            = &\sum_{\rho\in\mathrm{S}_{2K}} \te{Sgn}(\rho) \prod_{i=1}^{2n}\mathbf{C}_{i,\rho(i)} \left( \prod_{j=1}^n |T_{R_j,m}|^2 \right)^2
            \\
            = &\dfrac{1}{\mathcal{C}^2} \det(\mathbf{C}),
        \end{split}
    \end{equation}
    \bc{where $\mathcal{C} = \prod_{j=1}^n \gamma_j$. }
    Then we find that 
    \begin{equation}
        \begin{split}
            \det(\mathbf{C}) =& \left( \mathcal{C} + \mathcal{C}\mathcal{T}(\alpha_\te{d} - \beta_\te{d}) \right) \left( \mathcal{C} + \mathcal{C}\mathcal{T}(\alpha_\te{d} + \beta_\te{d}) \right),
        \end{split}
    \end{equation}
    where $\mathcal{T} = \sum_{j=1}^n 1/\gamma_j$.
    Consequently we have
    \begin{equation}
        \mathrm{det}(\mathbf{Q}_\mathcal{R}^{(m)}) =
        (1+\mathcal{T}\alpha_\te{d})^2 - (\mathcal{T}\beta_\te{d})^2.
    \end{equation}
    We then use this in Eq.(\ref{margnialp}) to obtain Eq.(\ref{tor_dis}).
    
\section*{Data Availability}

All data supporting the findings of this study are available upon request.

\section*{Code Availability}

The code used to generate the numerical results presented in this paper can be made available upon reasonable request.

\section*{Acknowledgments}
We acknowledge The Walrus Python library as a helpful reference on hafnian algorithms\cite{Gupt2019}.

This work is supported by the National Natural Science Foundation of China (62071301); State Council of the People’s Republic of China (D1210036A); NSFC Research Fund for International Young Scientists (11850410426); NYU-ECNU Institute of Physics at NYU Shanghai; the Science and Technology Commission of Shanghai Municipality (19XD1423000); the China Science and Technology Exchange Center (NGA-16-001); the NYU Shanghai Boost Fund.

\section*{Author Contributions}
    J.S. performed the calculations and developed the theory. T.B. conceived the work and supervised the project. Both authors wrote the paper.

\section*{Competing Interests}
    The authors declare no competing interests.

\section*{References}

\bibliography{z-ref_nature}

\end{document}